\documentclass[aps,%
twocolumn,
groupedaddress]{revtex4}
\usepackage{epsfig}

\begin{document}
\draft
\input epsf


\title{Tracing protons through the Galactic magnetic field: a clue for
  charge composition of ultra-high energy cosmic rays.}

\author{P.G.~Tinyakov$^{a,c}$ and I.I.~Tkachev$^{b,c}$\\
 $^a${\small\it Institute of Theoretical Physics, University of
 Lausanne,} \\ {\small\it CH-1015 Lausanne, Switzerland}\\
 $^b${\small\it CERN Theory  Division, CH-1211 Geneva 23, Switzerland}\\
 $^c${\small\it Institute for Nuclear Research, Moscow 117312,
 Russia } }

\begin{abstract}
We reconstruct the trajectories of ultra-high energy cosmic rays
(UHECR) --- observed by the AGASA experiment --- in the Galactic magnetic field
assuming that all particles have the same charge. We then study
correlations between the reconstructed events and BL Lacs. The
correlations have significance below $10^{-3}$ in the case of
particles with charge $+1$. In the case of charge $-1$ the
correlations are absent. We interpret this as evidence that protons
are present in the flux of UHECR. Observed correlation provides an
independent evidence that BL Lacs emit UHECR.
\end{abstract}


\maketitle

\section{Introduction}
Ultra-high energy cosmic rays (UHECR) are a subject of an active
debate for over more than 20 years. The data accumulated during this time
\cite{experiments,takeda} provide a compelling evidence that the GZK
cutoff \cite{GZK} predicted in the UHECR spectrum at energies $E\sim
10^{20}$~eV may be absent. Many models explaining this puzzle have
been proposed (for reviews see
Refs.~\cite{reviews}). The actual focus of the
debate is the question of whether an unconventional astrophysical
model can be constructed which explains the observed super-GZK events,
or a new physics is required. This key question still remains open.

With the accumulation of the high-energy events important signatures
emerge which discriminate efficiently between different models. It has
been known for some time that UHECR events form clusters
\cite{clusters1,takeda,Uchihori:2000gu}. Recent analysis shows that
clustering is statistically significant \cite{Tinyakov:2001ic}. The
small angular size of clusters of order $\simeq 2.5^{\circ}$, consistent
with the experimental angular resolution, suggests that they are due
to point sources. The models which do not reproduce this feature are
therefore disfavored.

Observation of clusters implies that some sources of UHECR may be
identified. Recently, significant correlations of arrival directions
of UHECR with positions of BL Lacertae (BL Lacs) were found
\cite{Tinyakov:2001nr}. BL Lacs are blazars (ANGs with relativistic
jet directed along the line of sight) characterized in particular
by the (near) absence of
the emission lines.  The correlations between UHECR and BL Lacs are
most significant at angles of order $\sim 2.5^{\circ}$ and are present
at relatively low energies $E > 2.4\times 10^{19}$. Such tightness
suggests strongly the existence of neutral primary particles.
Association with {\em distant} BL Lacs combined with {\em neutrality}
of primaries rules out most of the models of UHECR, leaving the models
based on neutrino \cite{Zbur} (the Z-burst models), models with hypothetical
``immune messengers'' \cite{hm}, or models involving
violation of the Lorentz invariance \cite{VLI}. It should be noted
that ultra-high energy photons also cannot be excluded at present
\cite{Kalashev:2001qp}.

Regardless of whether there exist neutral primary particles,
correlations with BL Lacs imply that the acceleration mechanism of UHECR
production actually works. Therefore, protons are involved and should
be present in the UHECR flux at energies around or below the GZK
cutoff. If identified, such protons provide independent evidence that
BL Lacs are indeed sources of UHECR. The purpose of this paper is to
address this issue.

Knowing actual sources of UHECR is an extremely powerful tool even
when statistics is limited. As we will see shortly, this tool can be
used to study charge composition of UHECR. The idea is to use the
bending of charged primary particles in the Galactic (GMF) \cite{GMF_bend}
and extragalactic (EGMF) \cite{EGMF_bend} magnetic fields. 
While deflections in EGMF are random and therefore
unpredictable, deflections in GMF are regular. If charges, energies of
particles and GMF are known, the original directions (before entering
GMF) can be restored. If the effect of EGMF is small, the actual
positions of sources can be reconstructed. 

In practice, one has to solve the inverse problem: to reconstruct
charges assuming set of potential sources. 
We show that this is possible at least in a
statistical sense: assuming that a substantial fraction of UHECR are
protons significantly improves correlations with BL Lacs. This {\em
implies } simultaneously that GMF model is roughly correct, and that
the effect of EGMF is small.

The paper is organized as follows. In Sect.~\ref{sect:GMF} we briefly
review the present knowledge about the Galactic magnetic field and fix
the GMF parameters for further calculations. In Sect.~\ref{sect:CA} we
analyse correlations between BL Lacs and UHECR when the latter are
assigned non-zero charges. Sect.~\ref{sect:conclusions} contains
discussion and concluding remarks.

\section{Galactic Magnetic Field}
\label{sect:GMF}

The Galactic magnetic field can be divided in two parts: the disc and the
halo. Each one has regular and turbulent components. While the strength of
turbulent component is larger, it is the regular field which gives dominant
contribution into CR propagation. Most of the information on the regular
component of the disc is obtained from the Faraday rotation measurements of
pulsars and extragalactic radio sources. The latter are used also for the
reconstruction of the halo field.  Magnetic field in the disc resembles the
spiral structure of our Galaxy. It may either reverse direction between
different spiral arms (bisymmetric, or BSS model), or there may be no
reversals (axisymmetric, or ASS model). Several field reversals were detected
\cite{RL,HQ,ID,HMQ,Frick} which are consistent with BSS model (note however
that discrimination between the two models is complicated by small-scale
irregularities in the magnetic field). In our calculations we adopt the BSS
model. Simple analytical representation of the spiral field structure (see
Refs. \cite{SF,HQ}) contains the following parameters:
\begin{itemize}
\item Distance from the Sun to the Galactic center, $R = 8.5$ kpc.
\item Local (at the Sun position in the Galaxy) field strength, $B_0$.
\item Pitch angle $p$ which determines the direction of the local magnetic
  field (the field points in the direction $l= 90+p$ in the Galactic
  coordinates).
\item Distance $d$ to the first field reversal. Negative $d$ means that
  nearest reversal occurs in the direction to the Galactic center, positive
  corresponds to the opposite direction.
\end{itemize}

In terms of these parameters the field in the disk is written as 
$$
B_\theta = B\, \cos(p), \qquad
B_r = B\, \sin(p) . 
$$
The magnitude $B=B(r,\theta)$ has the spiral structure,
\begin{equation}
B(r,\theta) = B(r)\, 
\, \cos\left(\theta - \beta \ln\left(\frac{r}{R}\right) + \phi \right) 
\, ,
\label{Bd}
\end{equation}
where $\beta \equiv 1/\tan(p)$, a constant phase $\phi$ is given by
\begin{equation}
\phi = \beta \ln\left(1 + \frac{d}{R}\right) - \frac{\pi}{2} \, ,
\end{equation}
and
\begin{equation}
B(r) = B_0 \;\frac{R}{r \cos (\phi )} \, .
\end{equation}
In the last expression the standard assumption that magnitude of the
field decreases as $r^{-1}$ in radial direction is made. It is also
assumed that $B(r) = {\rm const}$ at $r < 4$ kpc. Note that precise
dependence of the disc field far away from the Sun position is not
important for our study as only a small fraction of the observed UHECR
passes through this region. Note also that the constant phase
$\phi$ can be absorbed in another parameter $r_0$, so that
Eq. (\ref{Bd}) becomes $B(r,\theta) = B(r) \cos (\theta - \beta \ln
({r}/{r_0}) )$, which is the parametrization used in
Refs. \cite{SF,HQ}. We find it more convenient to work with
parameters directly related to the local field.

To proceed, we need to fix the parameters $B_0$, $p$ and $d$. All
studies of GMF based on pulsar and extragalactic rotation measures
converge on $B_0 = 1.4 \,\mu {\rm G}$, see \cite{RL,HQ,ID,Frick} and
recent reviews \cite{beck,han}. We adopt this value.

Pitch angle was found to be close to zero or positive in early
publications, $p=+5^{\circ}$ \cite{RK}, $p=-2^{\circ}$ \cite{RL}, but
decreased in more recent studies: $p=-8^{\circ}$ was obtained in
Refs.~\cite{HQ,Heiles,ID,HMQ}, while in Ref.~\cite{Frick}
$p=-15^{\circ}$ was found as an average pitch angle in nearby spiral
arms. Following reviews \cite{beck,han} we take $p=-8^{\circ}$.

Field reversal was found to be at $d = -0.2 - 0.3$~kpc in
Ref.~\cite{HQ} (however, the best fit value of $r_0$ in the same paper
corresponds to $d=-0.48$), at $d = -0.4$~kpc in Ref.~\cite{RL}, at $d
= -0.6$~kpc in Refs.~\cite{RK,Vale,Frick}. Finally, in the review
\cite{beck} $d = -0.6$~kpc was cited, while the review \cite{han}
follows Ref.~\cite{HQ}. We take $d = - 0.5$~kpc.

The simplest approximation for the halo field is obtained by taking
the disk field and extending it outside of the disk with exponentially
decreasing amplitude,
\begin{equation}
B(r,\theta,z) = \exp \left(-\frac{|z|}{h}\right)\ B(r,\theta) \; .
\label{Bt1}
\end{equation}
This introduces one more parameter, the height $h$.  Small values of
$B_z = \,0.2 \mu {\rm G}$ found in \cite{HQ} for the halo field
support this approximation.  The disc has a height of $h=1.5$~kpc
according to reviews \cite{beck,han} and we adopt this value.

In addition, the halo fields above and below the disk (more precisely,
its parts parallel to the disk) may be parallel or anti-parallel. 
In the latter case the halo field may be approximated by
\begin{equation}
B(r,\theta,z) = {\rm sign} (z)\, \exp \left(-\frac{|z|}{h}\right)\
B(r,\theta) \; .
\label{Bt}
\end{equation}
The first case, Eq. (\ref{Bt1}), corresponds to the quadrupole-type
model which we denote ${\rm BSS_Q}$, while the second case,
Eq. (\ref{Bt}), corresponds to the dipole-type model denoted as ${\rm
BSS_D}$ in what follows.  There are indications in favour of the ${\rm
BSS_D}$ global structure \cite{HMBB,beck,han} of the halo field.

In calculations of UHECR propagation in GMF, the parameters of the
latter are often chosen following early work \cite{Stanev:1996qj}.  We
do not use the conventions of Ref.~\cite{Stanev:1996qj} because they
are ambiguous: the parameter $r_0$ used there does not correspond to
the cited local field strength, the value of pitch angle does not
correspond to cited direction of the local field and references for
the assumed value of the halo height are not given.

To summarize, we adopt Eqs. (\ref{Bd}-\ref{Bt}) for the Galactic
magnetic field with the following set of parameters
\begin{eqnarray}
&& B_0 = 1.4 \, \mu {\rm G} \qquad p = -8^{\circ} \nonumber \\
&& d = -0.5 \, {\rm kpc} \qquad h = 1.5 \, {\rm kpc}
\label{GMF_par}
\end{eqnarray}
and assume that this field extends to $R_{\rm max} = 20 $ kpc in all
directions. It should be stressed that these parameters are chosen on
the basis of rotation measurements and their choice has nothing to do
with the propagation of UHECR.

\section{Correlation analysis}
\label{sect:CA}

We start by specifying the sets of BL Lacs and UHECR. Following
Ref.~\cite{Tinyakov:2001nr}, we take BL Lacs form the QSO catalog
\cite{veron} which contains 306 confirmed BL Lacs. The choice of UHECR
set is motivated as follows. Apriori, best results should be achieved
with the largest set having best angular and energy resolution. In
addition, this should be (relatively) high-energy set, because the
uncertainties in the deflection angle (which result from uncertainties
in GMF and energies of particles) should not dominate over the angular
resolution. This requirement is satisfied at
$E>4\times10^{19}$~eV. Therefore, we chose all published AGASA events
with energy $E>4\times 10^{19}$~eV \cite{takeda}. This set contains 57
events. 

For the quantitative measure of correlations we use the probability
$p(\delta)$ introduced in
Refs.~\cite{Tinyakov:2001ic,Tinyakov:2001nr}, which is a probability
to have certain excess of events within the angle $\delta$ from any of
the sources (BL Lacs). This probability is calculated by counting how
often the excess observed in the real data occurs in the Monte-Carlo
(MC) simulations. The MC configurations are generated as described in
Refs.~\cite{Tinyakov:2001ic,Tinyakov:2001nr}. The difference is that
now charges are assigned and arrival directions are corrected for GMF
both in the real data and in each MC set. The energies of MC events
are taken from the real data. Exactly the same treatment of the real
data and MC sets is important to prevent appearance of artificial
correlations.

The charge assignment can be done by {\em any} algorithm as long as
{\em the same} algorithm is used in MC simulations. In the simplest
case one assigns equal charges to all particles. Although correlations
which are due to neutral particles are destroyed in this case, this
effect may be compensated by charged particles which move closer to
their sources. One may expect this situation when fractions of neutral
and charged particles are comparable.

As explained in Ref.~\cite{Tinyakov:2001nr}, there are two possible
strategies to estimate the significance of correlations on the basis
of $p(\delta)$. One may impose cuts on BL Lac set and adjust them in
such a way that correlations are maximum. Likewise, one may look for
the "best" values of the parameters of GMF. In this case the penalty
factor should be calculated and included in the probability. Since
reconstruction of particle trajectories in GMF is rather
time-consuming, this approach is not very practical in the case at
hand. Thus, we do not adjust the parameters of GMF,
Eq.~(\ref{GMF_par}), and impose no cuts on BL Lac set except for a
single cut in the apparent magnitude. Instead of penalty calculation
we present explicit dependence of the probability $p(\delta)$ on this
cut. We draw our main conclusions from the behaviour of the curve,
rather than from the value of the probability at the minimum.
  
Correlations found in Ref.~\cite{Tinyakov:2001nr} were strongest at
$\delta \simeq 2.5^{\circ}$. We therefore fix this value of $\delta$
in our calculations. Since $\delta$ is not adjusted to minimize the
probability, there is no penalty factor associated with that.

\begin{figure}
\leavevmode\epsfxsize=3.25in\epsfbox{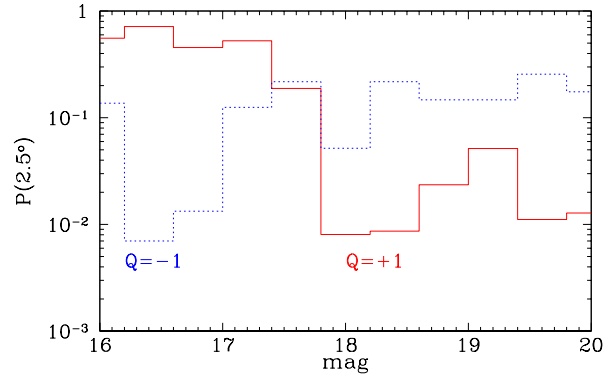}
\caption{The dependence of the probability $p(2.5^{\circ})$ on the 
  cut in magnitude in BL Lac catalog. Quadrupole-type GMF model, 
  Eq.~(\ref{Bt1}), is assumed.}
\label{BLL_Q}
\end{figure}

Consider first the case of the symmetric (quadrupole) model ${\rm BSS_Q}$.
Fig.~\ref{BLL_Q} shows the dependence of $p(2.5^{\circ})$ on the cut
in magnitude (the cut ${\rm mag}<20$ corresponds to inclusion of
practically all BL Lacs). Solid and dotted lines correspond to the
charge $+1$ or $-1$ assigned to all particles, respectively.
We see that in both cases $p(2.5^{\circ})$ has minima which are
comparable in depth (although both not very deep). One is tempted to
conclude that both charges may be present. This kind of situation is
expected in the Z-burst models. One may notice, however, that these
minima are not equally significant. The minimum at $Q=+1$ is wider and
corresponds to much higher statistics (correlations are present at the
level of $\sim 1\%$ even when all BL Lacs are included). Moreover,
event-by-event analysis shows that 12 out of 14 events contributing to
the minimum of probability at $Q=+1$ are situated in the Northern
hemisphere. This strange feature suggests that the field in the
Southern hemisphere is wrong, while in the Northern one it is roughly
correct. The obvious thing to try is the ${\rm BSS_D}$ model in which the
Southern field has different sign.

\begin{figure}
\leavevmode\epsfxsize=3.25in\epsfbox{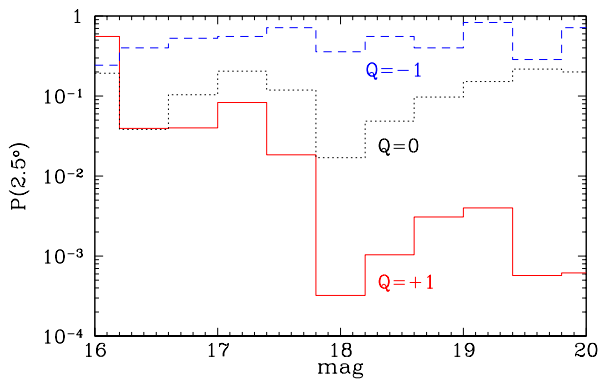}
\caption{The dependence of the probability $p(2.5^{\circ})$ on the 
  cut in magnitude in BL Lac catalog. Dipole-type GMF model, 
  Eq.~(\ref{Bt}), is assumed.}
\label{BLL_D}
\end{figure}

\begin{figure}
\leavevmode\epsfxsize=3.25in\epsfbox{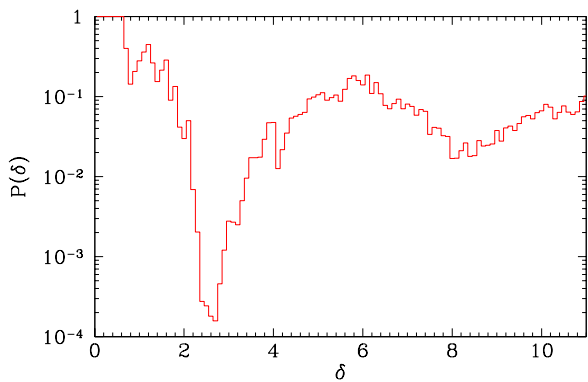}
\caption{The dependence of the probability $p(\delta)$ on the angle 
$\delta$ (in degrees) with the cut ${\rm mag} < 18$ in the BL Lac catalog
and charge $Q=+1$.
Dipole-type GMF model, Eq.~(\ref{Bt}), is assumed.}
\label{Pdelta}
\end{figure}

Consider the case of the asymmetric (dipole) model ${\rm BSS_D}$. In
our problem changing the direction of the GMF in the Southern
hemisphere to the opposite is equivalent to flipping sign of charges
of the events in the Southern hemisphere. According to previous
results, this should increase the correlations. Indeed, the situation
changes. Fig.~\ref{BLL_D} shows the dependence of the probability
$p(2.5^{\circ})$ on the cut in magnitude in the case of the ${\rm
BSS_D}$ model.
Three cases $Q=-1,0,+1$ are shown.  The correlations in the case
$Q=+1$ have improved by almost two orders of magnitude as compared to
the ${\rm BSS_Q}$ model and are now at the level below $10^{-3}$ in a wide
range of magnitudes. Even with no cuts on BL Lacs the significance of
correlations at $Q=+1$ is below $10^{-3}$. On the contrary, in the
case $Q=-1$ the correlations are now absent. In the case $Q=0$
correlations with BL Lacs satisfying ${\rm mag}<18$ are observed at
the level of $2\%$. For completeness, Fig.~\ref{Pdelta} shows the 
the dependence of $p(\delta)$ on the angle $\delta$ at the 
cut ${\rm mag} < 18$.

\section{Discussion and conclusions}
\label{sect:conclusions}

The pairs BL Lac -- cosmic ray separated by less than $2.5^{\circ}$
are listed in Table I (imposing the cut ${\rm mag} < 18$). 
BL Lacs appearing in this table are probable
sources of UHECR. For each BL Lac the name, Galactic coordinates l and
b, and redshift z (when known) are given. Corresponding cosmic ray
energy E (in units of $10^{19}$ eV) and probable charge Q are listed
in the last two columns.

We have assigned $Q=0$ to those particles which contribute to
correlations at $\delta =2.5^{\circ}$ in the case when all particles
are assumed to be neutral. Likewise, we have assigned charge $Q=+1$ to
those events which fall within $2.5^{\circ}$ from any of the BL Lacs
when they are assumed to have charge $+1$ and deflection in GMF is
taken into account.  Some events satisfy both requirements;
corresponding entry in Table I reads ``0 or +1''.

Lines 1-8 of Table~I contain four BL Lacs which correlate with two
cosmic rays each. These are most probable sources. In the lines 9-16
eight BL Lacs are listed which correlate with singlets.  Finally,
lines 17-22 contain three cosmic rays for which sources are ambiguous
(each ray has two neighbouring BL Lacs within $\delta <2.5^{\circ}$).

Examining Table~I one notices two striking regularities (we do not
attempt to assign statistical significance to those in the present
paper). First, most objects in this table are X-ray selected radio
loud BL Lacs. Second, the fraction of BL Lacs with unknown redshifts
in Table~I is much larger than in average over the whole BL Lac
catalog. In doublet part of the Table this fraction is 6:2, in singlet
section this fraction is 7:1. 

\begin{table}
\begin{tabular}{l|c|c|c|c|c|l}
& Name & l        &    b    &  z      &  \footnotesize E &  \footnotesize Q\\ 
\hline
\footnotesize 1 &\footnotesize 2EG J0432+2910    &\footnotesize 170.52 &\footnotesize -12.6  %
&\footnotesize -  &\footnotesize 5.47  &\footnotesize 0 or +1\\
\footnotesize 2 &\footnotesize     &\footnotesize  &\footnotesize   %
&\footnotesize   &\footnotesize 4.89  &\footnotesize 0 or +1\\
\footnotesize 3 &\footnotesize RX J1838.7+4802    &\footnotesize 76.95 &\footnotesize 21.83  %
&\footnotesize -  &\footnotesize 10.6 &\footnotesize 0 or +1 \\
\footnotesize 4 &\footnotesize     &\footnotesize  &\footnotesize   %
&\footnotesize   &\footnotesize 4.35 &\footnotesize +1 \\
\footnotesize 5 &\footnotesize RGB J0109+182    &\footnotesize 128.82 &\footnotesize -44.4  %
&\footnotesize -  &\footnotesize 21.3  &\footnotesize +1\\
\footnotesize 6 &\footnotesize     &\footnotesize  &\footnotesize   %
&\footnotesize   &\footnotesize 5.07  &\footnotesize 0 or +1\\
\footnotesize 7 &\footnotesize RX J1058.6+5628    &\footnotesize 149.59 &\footnotesize 54.42  %
&\footnotesize 0.144  &\footnotesize 7.76  &\footnotesize 0\\
\footnotesize 8 &\footnotesize     &\footnotesize  &\footnotesize   %
&\footnotesize   &\footnotesize 5.35  &\footnotesize 0\\
\hline
\footnotesize 9 &\footnotesize RGB J1415+485    &\footnotesize 91.2 &\footnotesize  63.11 %
&\footnotesize  - &\footnotesize 6.22  &\footnotesize +1\\
\footnotesize 10 &\footnotesize RX J0035.2+1515 &\footnotesize 117.15 &\footnotesize -47.44  %
&\footnotesize -  &\footnotesize 5.53  &\footnotesize 0 or +1\\
\footnotesize 11 &\footnotesize RX J1704.8+7138    &\footnotesize 103.09 &\footnotesize 33.96  %
&\footnotesize -  &\footnotesize 4.78  &\footnotesize +1\\
\footnotesize 12 &\footnotesize  OT 465   &\footnotesize 74.22 &\footnotesize  31.4 %
&\footnotesize -  &\footnotesize 4.88  &\footnotesize 0 or +1\\
\footnotesize 13 &\footnotesize RX J1702.6+3115    &\footnotesize 53.4 &\footnotesize 35.76  %
&\footnotesize -  &\footnotesize 4.47  &\footnotesize 0 or +1\\
\footnotesize 14 &\footnotesize RX J1359.8+5911    &\footnotesize 107.36 &\footnotesize 55.83  %
&\footnotesize -  &\footnotesize 4.46  &\footnotesize 0 \\
\footnotesize 15 &\footnotesize RGB J0159+107    &\footnotesize 148.75 &\footnotesize -48.64  %
&\footnotesize -  &\footnotesize 4.2  &\footnotesize +1\\
\footnotesize 16 &\footnotesize 1ES 1853+671    &\footnotesize 97.74 &\footnotesize 24.63  %
&\footnotesize 0.212  &\footnotesize 4.39  &\footnotesize +1\\
\hline
\footnotesize 17 &\footnotesize RX J1100.3+4019    &\footnotesize 175.87 &\footnotesize 63.56  %
&\footnotesize -  &\footnotesize 7.21  &\footnotesize 0\\
\footnotesize 18 &\footnotesize EXO 1118.0+4228    &\footnotesize 167.85 &\footnotesize 66.16  %
&\footnotesize 0.124  &\footnotesize   &\footnotesize 0 or +1\\
\hline
\footnotesize 19 &\footnotesize RGB J1426+340    &\footnotesize 57.6 &\footnotesize 68.53  %
&\footnotesize -  &\footnotesize 4.97  &\footnotesize +1\\
\footnotesize 20 &\footnotesize TEX 1428+370    &\footnotesize 63.95 &\footnotesize 66.92  %
&\footnotesize 0.564  &\footnotesize   &\footnotesize +1\\
\hline
\footnotesize 21 &\footnotesize B2 0804+35    &\footnotesize 186.48 &\footnotesize 30.35  %
&\footnotesize 0.082  &\footnotesize 4.09  &\footnotesize +1 \\
\footnotesize 22 &\footnotesize TXS 0806+315    &\footnotesize 190.42 &\footnotesize 29.36  %
&\footnotesize 0.22  &\footnotesize   &\footnotesize +1\\
\end{tabular}
\caption{The list of pairs BL Lac -- cosmic ray which
contribute to correlations of Fig.~\ref{BLL_D} at $\mbox{mag}<18$.}
\end{table}


Absence of emission lines (more precisely, their weakness and
narrow width) is a defining feature of BL Lac family within the
general blazar class. Therefore, it is not surprising that redshifts
of roughly half of confirmed BL Lacs are not known. Increased fraction
of such BL Lacs in Table~I may mean that the absence of emission lines is
important for a blazar to became emitter of UHECR.  

As it is generally assumed, BL Lacs with unknown redshift may be far
away: the observational bias would then explain that the redshifts of
such objects are unknown more often. Then however new physics or
extreme astrophysics would be necessary to explain observed
correlations.

Within the conventional framework, when propagation is limited by
interactions with comic backgrounds, one should expect sources to be
relatively nearby. Closest BL Lac with known redshift is at $z=0.029$.
This is well outside of the GZK sphere. It should be noted, however,
that most of the cosmic rays in Table I have energies below the GZK
cutoff. As it was found in Ref.~\cite{Kalashev:2001qp}, the flux of
protons at energy $E = 8\times 10^{19}$ is not attenuated
substantially if sources are located at $z<0.03$, and the flux at $E =
5\times 10^{19}$ is not attenuated if sources are located at $z<0.1$.
Assuming this range of redshifts for charged entries in Table~I, we
see that there is no problem to explain all of them except lines 3 and
5. The ray (21-22) has sufficiently low energy so that BL Lac B2
0804+35 with z=0.082 can be a source without any problem with
attenuation.  Likewise, the BL Lac 19 can be an actual source of the
ray (19-20) if it has redshift $z \alt 0.1$. Thus, the correlations
due to charged particles listed in Table~I can be explained within the
framework of conventional physics if sources with unknown redshifts
are assumed to be within $z \alt 0.1$.

Consider now the entries corresponding to neutral particles. According
to Ref.~\cite{Kalashev:2001qp}, photons can be UHECR primaries at any
energy $E >10^{19}$~eV provided that EGMF is smaller than $10^{-9}$~G,
energy spectrum at the source is hard, $\propto E^{-\alpha}$ with
$\alpha < 2$ and the maximum energy is high enough.  (Note that
correlations with charged particles imply that EGMF is in this range
anyway.) Therefore, entries 3 and 17 can be explained by photons.  The
real difficulty is with rays 5,7,8 and 16. Note, however, that
according to MC simulations, there should be 7 background events in
the Table~I on average.

To summarize, the idea of using the Galactic magnetic field as a
mass-spectrograph of UHECR seems to work. The correlations between
UHECR and BL Lacs substantially improve when arrival directions of
cosmic rays are corrected for GMF. If not a statistical fluctuation,
this implies the following: i)~cosmic rays of highest energies contain
a substantial fraction of protons ii)~extragalactic magnetic fields
have little effect on propagation of UHECR even from cosmological
distances iii)~the model of the Galactic magnetic field
described by Eq.~(\ref{Bt}) is roughly
correct. 

Finally, the significance of correlations with charged
particles is $p < 10^{-3}$. This may be considered as yet more
evidence that BL Lacs are sources of UHECR. Interestingly, the events
and BL Lacs contributing into the correlation with lowest $p$ 
found in Ref.~\cite{Tinyakov:2001nr}
and the charged events of Table~I do not overlap, so the two
correlations should be considered as independent.

\section*{Acknowledgments}
{\tolerance=400 We are grateful to J. L. Han, M. E. Shaposhnikov,
D. V. Semikoz and P. Veron for valuable comments and discussions.  The
work is supported by the Swiss Science Foundation, grant 21-58947.99
and by INTAS grant 99-1065.}

\end{document}